\def\Pb{P_{\rm b}}

\def\rfr#1{Equation\,(\ref{#1})}
\def\rfrs#1#2{Equations\,(\ref{#1})-(\ref{#2})}

\def\virg#1{``#1"}

\def\eqi{\begin{equation}}
\def\eqf{\end{equation}}
\def\eqia{\begin{eqnarray}}
\def\eqfa{\end{eqnarray}}

\def\rp#1#2{{#1\over#2}}
\def\lb#1{\label{#1}}

\def\bds#1{\boldsymbol{#1}}

\def\ton#1{\left(#1\right)}

\def\grf#1{\left\{#1\right\}}

\documentclass[linenumbers]{aastex}

\usepackage{morefloats}
\usepackage[title]{appendix}
\usepackage{hyperref,textcomp}
\usepackage{booktabs}
\usepackage[table,xcdraw]{xcolor}
\usepackage{multirow}
\usepackage{rotating,tabularx}
\usepackage{float}
\usepackage{enumerate}
\usepackage{rotating}
\usepackage[polutonikogreek,english]{babel}
\usepackage{amsmath,starfont,textgreek,w-greek,wasysym}
\usepackage[flushleft]{threeparttable}
\usepackage{amsthm}
\usepackage{amscd,lineno}
\usepackage{amssymb,dsfont}
\usepackage{graphicx,epsfig}
\usepackage[utf8]{inputenc}
\usepackage[T1]{fontenc}
\usepackage{txfonts}
\usepackage{xr-hyper}

\RequirePackage{color}

\newcommand{\emaila}{lorenzo.iorio@libero.it}

\allowdisplaybreaks[1]

\begin{document}

\title{The Impact of Classical and General Relativistic Obliquity Precessions on the Habitability of Circumstellar Neutron Stars' Planets}

\shortauthors{L. Iorio}

\author{Lorenzo Iorio\altaffilmark{1} }
\affil{Ministero dell'Istruzione, dell'Universit\`{a} e della Ricerca
(M.I.U.R.)
\\ Viale Unit\`{a} di Italia 68, I-70125, Bari (BA),
Italy}

\email{\emaila}

\begin{abstract}
Recently, it has been shown that rocky planets orbiting neutron stars can be habitable under not unrealistic circumstances. If a distant, point-like source of visible light, such as a Sun-like main-sequence star or the gravitationally-lensed accretion disk of a supermassive black hole is present, possible temporal variations $\Delta\varepsilon_\mathrm{p}\ton{t}$ of the planet's axial tilt $\varepsilon_\mathrm{p}$ to the ecliptic plane should be included in the overall habitability budget since the obliquity determines the insolation at a given latitude on a body' s surface. I point out that, for rather generic initial spin-orbit initial configurations, general relativistic and classical spin variations induced by the post-Newtonian de Sitter and Lense--Thirring components of the field of the host neutron star and by its pull to the planetary oblateness $J_2^\mathrm{p}$ may induce huge and very fast variations of $\varepsilon_\mathrm{p}$ that would likely have an impact on the habitability of such worlds. In particular, for a planet's distance of, say, $0.005\,\mathrm{au}$ from a $1.4\,M_\odot$ neutron star corresponding to an orbital period $\Pb=0.109\,\mathrm{day}$, obliquity shifts $\Delta\varepsilon_\mathrm{p}$ as large as $\varepsilon^\mathrm{max}_\mathrm{p}-\varepsilon_\mathrm{p}^\mathrm{min}\simeq 50^\circ-100^\circ$ over characteristic timescales as short as $10\,\mathrm{days}$ ($J_2^\mathrm{p}$) to $3\,\mathrm{Myr}$ (Lense--Thirring) may occur for arbitrary orientations of the orbital and spin angular momenta $\bds L,\,{\bds S}_\mathrm{ns},\,{\bds S}_\mathrm{p}$ of the planet-neutron star system. In view of this feature of their spins, I dub such hypothetical planets as \virg{nethotrons.}
\end{abstract}

{
\textit{Unified Astronomy Thesaurus concepts}: Exoplanets\,(498);  Neutron stars(1108); General relativity\,(641); Oblateness(1143); Astrobiology\,(74); Gravitation\,(661); 
}

\section{Introduction}
I will consider a novel type of extrasolar habitat potentially capable to sustain life whose existence, although still speculative, may not be deemed as  unrealistic. It is a restricted two-body system $\mathcal{S}$ made of an Earth (or, better, a super-Earth)-like planet (p) in a relatively close orbit about a neutron star ns. In turn, $\mathcal{S}$ orbits an external, distant point-like source S of electromagnetic radiation, whose distance and luminosity can be adjusted in order to have an overall p's irradiation from both S and ns compatible with life on it. In particular, S should provide the required amount of visible light lacking from the output of ns that, in combination with the characteristics of the planetary atmosphere, would essentially assure the presence of  liquid water and a moderate temperature on p for a sufficiently long time \citep{2017A&A...608A.147P}.  I propose to name such a putative planet  \virg{nethotron,} from the union of the verb $\upnu\acute{\upeta}\uptheta\upomega$ ($n\acute{\bar{e}}th\bar{o}$), meaning \virg{I spin}, and of the suffix --$\uptau\uprho\upo\upnu$ (--$tron$), which forms instrument nouns. My suggestion is due to the fact that, as I will show, the tilt $\varepsilon_\mathrm{p}$ of the planetary spin angular momentum ${\bds S}_\mathrm{p}$ to the  plane of the orbital motion of $\mathcal{S}$ about S--defined from
\eqi
\cos\varepsilon_\mathrm{p}={\bds{\hat{S}}}_\mathrm{p}\bds\cdot\bds{\hat{W}},
\eqf
where $\bds W$ is the orbital angular momentum of $\mathcal{S}$ around S--undergoes huge and fast temporal variations due, among other things, also to the General Theory of Relativity  \citep{1975PhRvD..12..329B,2014grav.book.....P}.  They are independent of the certain peculiar p's own physical and orbital properties like radius $R_\mathrm{p}$, quadrupole mass moment $J_2^\mathrm{p}$, spin $S_\mathrm{p}$, etc., which, in turn, induce further, generally faster changes in $\varepsilon_\mathrm{p}$ \citep{1975PhRvD..12..329B,2011CeMDA.111..105C}. Such a distinctive feature of $\varepsilon_\mathrm{p}$ can be quite relevant for life.

Actually, a handful of rocky planets in relatively close orbits, $\simeq 0.2-0.5\,\mathrm{au}$ wide, were discovered around the millisecond pulsar PSR B1257+12 \citep{1992Natur.355..145W,1994Sci...264..538W}, while a Jupiter-sized chtonian\footnote{For such a definition, see \citet{2004ASPC..321..203H}.} body made of carbon and oxygen was found orbiting the millisecond pulsar PSR J1719-1438 at just $0.004\,\mathrm{au}$ from it \citep{2011Sci...333.1717B}. In fact, according to \citet{2017A&A...608A.147P}, most neutron stars may host a wide variety of planetary systems, not yet discovered mainly because of observational biases. It was recently shown \citep{2017A&A...608A.147P} that pulsar planets, despite the fierce radiation environment in which they are immersed, can be habitable under certain, not unrealistic, assumptions about the mass of their atmosphere, the presence or not of a magnetic field, the X-ray irradiation, and the relativistic wind from the pulsar. In particular, the two super Earths orbiting PSR B1257+12 may be habitable, depending on still poorly constrained aspects of the pulsar wind \citep{2017A&A...608A.147P}. A neutron star habitable zone, based on the presence of liquid water and retention of an atmosphere, was defined by \citet{2017A&A...608A.147P} and depicted in their Figure\,5, as an allowed region in a plot having the pulsar's luminosity (X-ray plus wind) on the vertical axis and the orbital distance, ranging from $0.001--10\,\mathrm{au}$ on the horizontal axis. It should be stressed that the findings by \citet{2017A&A...608A.147P} were obtained for an isolated host neutron star; thus, they should likely be revised in presence of a possible external source S of solar-type irradiation at the right distance.

As far as the latter  is concerned, it could be, e.g., an ordinary Sun-like main-sequence star at, say, about $1\,\mathrm{au}$ from $\mathcal{S}$ or so; I dub such a scenario a circumstellar nethotron (CSN). By noting that the effects of the star's irradiation would sum up with those by the host neutron star itself, it may turn out that its spectral class and distance  might be different from those of the Sun--Earth case. Despite the companions of the binary pulsars hosting main-sequence stars \citep{1998MNRAS.298...67W}, what have been discovered so far  are much more massive than the Sun and
it is plausible that binaries made of a neutron star and a Sun-type star exist, e.g., in globular clusters due to exchange interactions in the core and in the Galactic field as regular outcomes of the evolution of binaries made of an initially massive star ($M_\star\gtrsim 8-10\,M_\odot$) and a Sun-like one. The number of binaries made of a neutron stars and  a Sun-like main-sequence star should be easily evaluated by population synthesis\footnote{Postnov (2021, private communication),}.

On the other hand, S could even be the lensed image of the accretion disk around a supermassive black hole (SMBH) at such a  distance that it would look like the Sun as viewed from the Earth \citep{2019arXiv191000940S}. In this context, I use the denomination of a circumnigricon nethotron (CNN), from the adjective \virg{\textit{niger}} meaning \virg{black.} It should be recalled that the existence of a large number of neutron stars in the neighborhood of, e.g., the SMBH in $\mathrm{Sgr\,A}^\ast$ at the Galactic Center,  was hypothesized long ago \citep{1996ARA&A..34..645M,2010RvMP...82.3121G,2018JKAS...51..165K}, and their search is currently actively pursued \citep{1997ApJ...475..557C,2017AAS...22943103C,2017MNRAS.471..730R,2019BAAS...51c.438B}. Moreover, it was recently shown  that a large number of (starless) potentially habitable Earth-sized planets, dubbed \virg{blanets} by their proponents, can form in SMBHs' accretion disks \citep{2019ApJ...886..107W,2021ApJ...909...96W,2021ApJ...907...58W}.

In summary, a CSN or CNN, in addition to the energy output from its host neutron star, would receive  an additional amount of electromagnetic radiation from S, which may be comparable, to some extent, to that of the Earth from the Sun. Among the several parameters that pose constraints on the long-term habitability of a world, the obliquity $ \varepsilon_\mathrm{p}$ is one of the most important ones since it determines the irradiation  on the planetary surface at a given latitude \citep{2009ApJ...691..596S}. Large and fast temporal variations
\eqi
\Delta \varepsilon_\mathrm{p}\doteq \varepsilon_\mathrm{p}(t)- \varepsilon_\mathrm{p}^0
\eqf
of $ \varepsilon_\mathrm{p}$ with respect to some reference value $ \varepsilon_\mathrm{p}^0$ at the epoch are generally believed to be detrimental to sustaining life and civilizations; although, under certain circumstances, they can help in avoiding snowball states by pushing the outer limit of the habitable zone outward; see, e.g., \citet{2014AsBio..14..277A} and \citet{2018AJ....155..266D}. Be that as it may, such an effect must be taken into account in establishing the overall habitability conditions on a planet. I will show that the spin axis ${\bds{\hat{S}}}_\mathrm{p}$ of a nethotron may experience wild variations $\Delta \varepsilon_\mathrm{p}$ with respect to the ecliptic plane of $\mathcal{S}$ over very short timescales directly induced, among other things, by the post-Newtonian (pN) gravitoelectric (Schwarzschild) and gravitomagnetic (Lense--Thirring) components of the field of ns. Depending on the size of the quadrupole mass moment $J_2^\mathrm{p}$ of the nethotron, a further spin rate of classical origin is present as well. The resulting variations are, in general, much faster than the pN ones and do not cancel them.

The paper is organized as follows. In Section\,\ref{seconda}, I present the model that I use in Section\,\ref{terza} for numerically calculating the shifts $\Delta\varepsilon_\mathrm{p}\ton{t}$ by varying the initial spin-orbit configuration. Section\,\ref{fine} summarizes my findings and offers my conclusions.
\section{The Spin-Precession Model}\lb{seconda}
Here, I described the coupled averaged equations in vectorial form for the rates of change of the spin angular momenta ${\bds S}_\mathrm{p},\,{\bds S}_\mathrm{ns}$ of p and ns, respectively; of the orbital angular momenta $\bds L,\,\bds W$ of the astrocentric motion about ns and of $\mathcal{S}$ about S, respectively; and of the Laplace-Runge-Lenz vector $\bds e$ of the planet-neutron star orbital motion within $\mathcal{S}$ retrieved from \citet{1975PhRvD..12..329B} and \citet{2011CeMDA.111..105C}.
In particular, I included
\begin{enumerate}
  \item The pN two-body de Sitter rate of the spin angular momentum ${\bds S}_\mathrm{p}$ of p due to the masses $M_\mathrm{ns}$ and $M_\mathrm{p}$ of ns and p \citep[][Equation\,(44) with Equation\,(45)]{1975PhRvD..12..329B}
  \item The pN Lense--Thirring rate of the spin angular momentum ${\bds S}_\mathrm{p}$ of p due to the spin angular momentum $\bds S_\mathrm{ns}$ of ns \citep[][Equation\,(44) with Equation\,(46)]{1975PhRvD..12..329B}
  \item The Newtonian rate of the spin angular momentum ${\bds S}_\mathrm{p}$ of p due to the pull of the mass $M_\mathrm{ns}$ of ns on the quadrupole mass moment $J_2^\mathrm{p}$ of p \citep[][Equation\,(44) with Equation\,(47)]{1975PhRvD..12..329B} or \citep[][Equations\,(10)-(11)]{2011CeMDA.111..105C}
  \item The Newtonian rate of the spin angular momentum ${\bds S}_\mathrm{p}$ of p due to the pull of the distant mass $M_\mathrm{S}$ of S on the quadrupole mass moment $J_2^\mathrm{p}$ of p \citep[][Equation\,(10) with Equation\,(12)]{2011CeMDA.111..105C}
  \item The Newtonian two-body rate of the orbital angular momentum $\bds L$ due to the quadrupole mass moment  $J_2^\mathrm{ns}$ of ns \citep[][Equation\,(64) with Equation\,(72)]{1975PhRvD..12..329B}
  \item The Newtonian two-body rate of the orbital angular momentum $\bds L$ due to the quadrupole mass moment $J_2^\mathrm{p}$ of p \citep[][Equation\,(64) with Equation\,(71)]{1975PhRvD..12..329B}
  \item The Newtonian rate of the orbital angular momentum $\bds L$ due to the torque by the distant mass $M_\mathrm{S}$ of S \citep[][Equation\,(7) with Equation(13)]{2011CeMDA.111..105C}
  \item The pN two-body Lense--Thirring rate of the orbital angular momentum $\bds L$ due to the spin angular momentum $\bds S_\mathrm{ns}$ of ns \citep[][Equation\,(64) with Equation\,(69)]{1975PhRvD..12..329B}
  \item The pN two-body Lense--Thirring rate of the orbital angular momentum $\bds L$ due to the spin angular momentum ${\bds S}_\mathrm{p}$ of p \citep[][Equation\,(64) with Equation\,(68)]{1975PhRvD..12..329B}
  \item The pN two-body rate \`{a} la Stern-Gerlach of the orbital angular momentum $\bds L$ due to both the spin angular momenta ${\bds S}_\mathrm{p},\,\bds S_\mathrm{ns}$ of p and ns \citep[][Equation\,(64) with Equation\,(70)]{1975PhRvD..12..329B}
  \item The pN two-body de Sitter rate of the spin angular momentum $\bds S_\mathrm{ns}$ of ns due to the masses $M_\mathrm{ns}$ and $M_\mathrm{p}$ of ns and p  \citep[][Equation\,(44) with Equation\,(45) and (1)$\leftrightarrows$(2)]{1975PhRvD..12..329B}
  \item The pN Lense--Thirring rate of the spin angular momentum $\bds S_\mathrm{ns}$ of ns due to the spin angular momentum ${\bds S}_\mathrm{p}$ of p  \citep[][Equation\,(44) with Equation\,(46) and (1)$\leftrightarrows$(2)]{1975PhRvD..12..329B}
  \item The Newtonian rate of the spin angular momentum $\bds S_\mathrm{ns}$ of ns due to the pull of the mass $M_\mathrm{p}$ of p on the quadrupole mass moment $J_2^\mathrm{ns}$ of ns  \citep[][Equation\,(44) with Equation\,(47) and (1)$\leftrightarrows$(2)]{1975PhRvD..12..329B} or \citep[][Equations\,(10)-(11)]{2011CeMDA.111..105C}
  \item The Newtonian rate of the spin angular momentum ${\bds S}_\mathrm{ns}$ of ns due to the pull of the distant mass $M_\mathrm{S}$ of S on the quadrupole mass moment $J_2^\mathrm{ns}$ of ns \citep[][Equation\,(10) with Equation\,(12)]{2011CeMDA.111..105C}
  \item The Newtonian rate of the orbital angular momentum $\bds W$ of $\mathcal{S}$ due to the torques exerted by the distant mass $M_\mathrm{S}$ coupled to the masses $M_\mathrm{p},\,M_\mathrm{ns}$ of p and ns and to their quadrupole mass moments as well $J_2^\mathrm{p},\,J_2^\mathrm{s}$ \citep[][Equation\,(8) with Equations(12)-(13)]{2011CeMDA.111..105C}.
  \item The Newtonian rate of the Laplace--Runge--Lenz vector $\bds e$ of the astrocentric orbital motion of p due to the torques exerted by the distant mass $M_\mathrm{S}$ and by the quadrupole mass moments $J_2^\mathrm{p},\,J_2^\mathrm{s}$ of p and ns \citep[][Equation\,(9) with Equation(11) and Equation(13)]{2011CeMDA.111..105C}.
  \item  The pN rate of the Laplace--Runge--Lenz vector $\bds e$ of the astrocentric orbital motion of p due to the gravitoelectric and gravitomagnetic pN components of the fields of p and ns  \citep[][Equation\,(65) with Equations(67)-(70)]{1975PhRvD..12..329B}
\end{enumerate}
For the sake of a meaningful comparison with \citet{1975PhRvD..12..329B}, the body (1) is identified with, say, p and the body (2) is ns. Moreover, the dimensionful quadrupolar parameter $\Delta I^{\ton{j}},\,j=1,\,2$ of \citet{1975PhRvD..12..329B} corresponds to $J^j_2\,M_j\,R_j^2$ for p and ns, respectively. Thus, the correspondences ${\vec{\mathrm{n}}}^{(1)}\rightarrow{\bds{\hat{S}}}_\mathrm{p},\,{\vec{\mathrm{n}}}^{(2)}\rightarrow{\bds{\hat{S}}}_\mathrm{ns},\,
{\vec{\mathrm{n}}}\rightarrow{\bds{\hat{L}}}$ hold. As far as \citet{2011CeMDA.111..105C} is concerned, the correspondences for vectors and versors are
${\bds G}_1\rightarrow\bds L$,
${\bds G}_2\rightarrow\bds W$,
${\bds L}_0\rightarrow{\bds S}_\mathrm{ns}$,
${\bds L}_1\rightarrow{\bds S}_\mathrm{p}$,
${\bds{\hat{k}}}_1\rightarrow\bds{\hat{L}}$,
${\bds{\hat{k}}}_2\rightarrow\bds{\hat{W}}$,
${\bds{\hat{s}}}_0\rightarrow{\bds{\hat{S}}}_\mathrm{ns}$,
and  ${\bds{\hat{s}}}_1 \rightarrow{\bds{\hat{S}}}_\mathrm{p}$. Furthermore, the correspondences for the suffixes of masses, radii, quadrupole mass moments, etc., are $0\rightarrow\mathrm{ns},\,1\rightarrow\mathrm{p},\,2\rightarrow\mathrm{S}$.
\section{Numerically Produced Time Series for $\Delta\varepsilon_\mathrm{p}$}\lb{terza}
Here, I numerically integrate the coupled equations of Section\,\ref{seconda} for chosen values of the relevant physical and orbital parameters of p, ns, and S in, say, a CSN scenario and by varying the initial spin-orbit configuration in order to account for possible different formation and evolution channels of nethotrons. The goal of my analysis is not designing any accurate test of pN gravity; instead, it aims at preliminarily investigating if obliquity variations, whose amplitudes may be significant for the nethotron's ability to sustain life, occur within relatively short timescales for rather general initial spin-orbit configurations. Indeed, in view of our currently poor and debated knowledge of the formation and evolution channels of planets around neutron stars \citep{1995ASPC...72..411P,2000MNRAS.316L..21G,2006ISSIR...6...77W,2014P&SS..100...19W}, it would be unnecessarily restrictive to limit ourselves  to some specific orientation of the three angular momenta within $\mathcal{S}$.

I adopt a prototypical neutron star with  mass and equatorial radius $M_\mathrm{ns}=1.4\,M_\odot,\,R_\mathrm{ns}=11\,\mathrm{km}$, respectively \citep{2021PhRvL.126r1101S}. For its spin frequency $\nu_\mathrm{ns}$, dimensional rotational quadrupole moment $Q_\mathrm{ns}$ and moment of inertia $I_\mathrm{ns}$, I rely upon the recently measured values for the isolated millisecond pulsar PSR J0030+0451 \citep{2021PhRvL.126r1101S}.
The (negative) dimensional quadrupole mass moment $Q_\mathrm{ns}$ is related to the dimensionless one by
\eqi
J^\mathrm{ns}_2 = -\rp{Q_\mathrm{ns}}{M_\mathrm{ns}\,R_\mathrm{ns}^2}.
\eqf
For its value, following the recent results by \citet{2021PhRvL.126r1101S}, I assume
\eqi
\left|Q_\mathrm{ns}\right| \simeq 1.5\times 10^{36}\,\mathrm{kg\,m}^2,\lb{Qu2}
\eqf
while for the neutron star's moment of inertia $I_\mathrm{ns}$ I use \citep{2021PhRvL.126r1101S}
\eqi
I_\mathrm{ns} \simeq 1.6\times 10^{38}\,\mathrm{kg\,m}^2.
\eqf
For, say \citep{2021PhRvL.126r1101S}
\eqi
\nu_\mathrm{ns}=200\,\mathrm{Hz}
\eqf
corresponding to a spinning period of about
\eqi
P_\mathrm{ns}\simeq 5\,\mathrm{ms},
\eqf
the neutron star's spin angular momentum is
\eqi
S_\mathrm{ns} = I_\mathrm{ns}\,2\uppi\nu_\mathrm{ns} \simeq 2\times 10^{41}\,\mathrm{J\,s}.\lb{spinNS}
\eqf

About the nethotron, it is likely that it should be more massive than our planet in order to keep a sufficiently huge atmosphere in order to mitigate the erosion caused by the harsh radiation from the host neutron star \citep{2017A&A...608A.147P}. Given the present uncertainties in the characteristic physical parameters of super-Earth planets \citep{2013AREPS..41..469H,2015enas.book.2436H}, for the sake of simplicity, I model the nethotron as a body with the same mass $M_\mathrm{p}$, radius $R_\mathrm{p}$, dimensionless quadrupole mass moment $J_2^\mathrm{p}$ and spin angular momentum $S_\mathrm{p}$ of Earth.

I assume a Sun-like main-sequence star at 1 au from $\mathcal{S}$ as distant source S of visible light.

\begin{figure}[H]
\centering
\centerline{
\vbox{
\begin{tabular}{c}
\epsfxsize= 16 cm\epsfbox{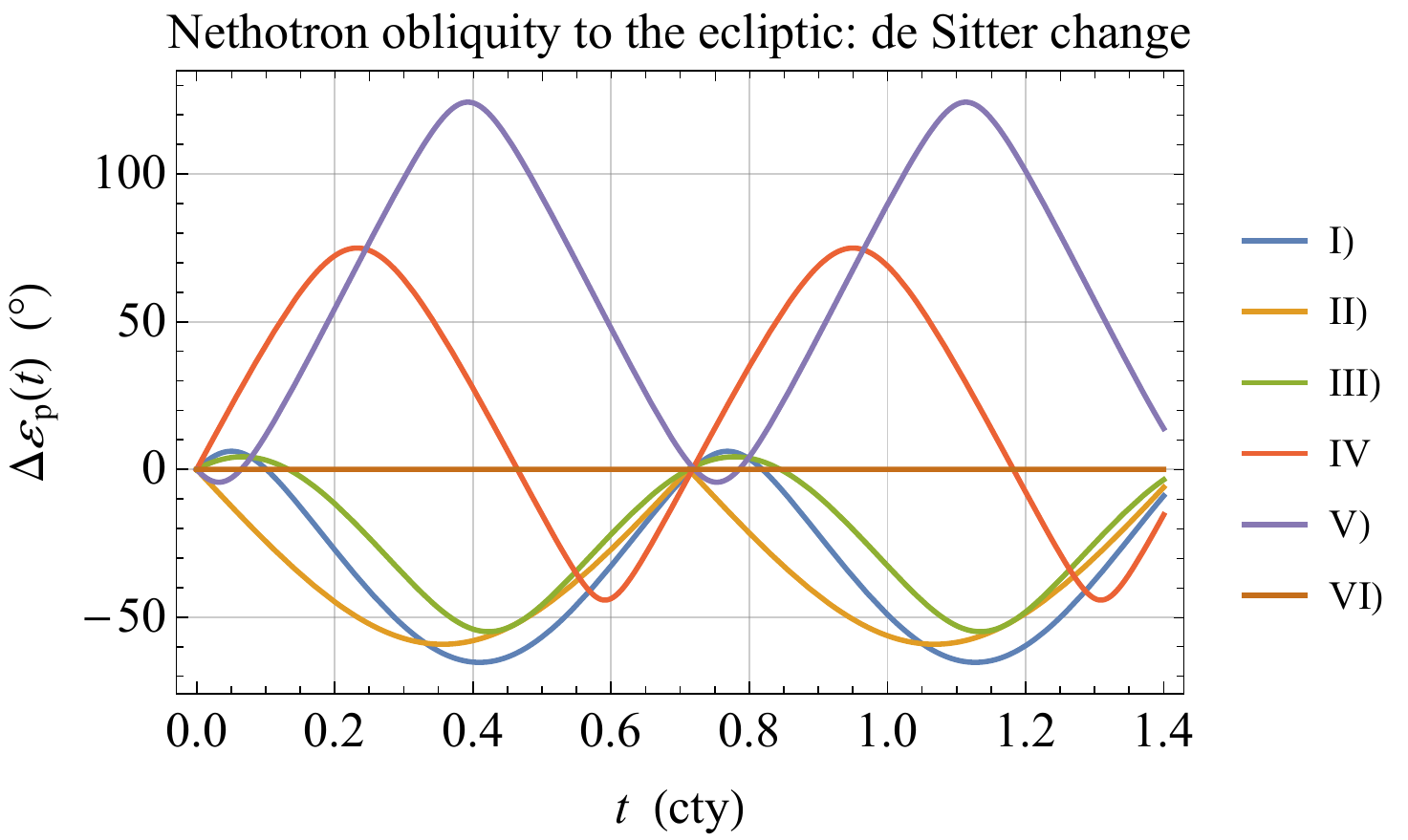}\\
\end{tabular}
}
}
\caption{
Numerically produced time series $\Delta \varepsilon_\mathrm{p}\ton{t} =  \varepsilon_\mathrm{p}\ton{t}- \varepsilon_\mathrm{p}^0$, in $^\circ$, of the pN gravitoelectric variation of the obliquity $ \varepsilon_\mathrm{p}$ to the ecliptic plane of a putative Earth-like netohotron orbiting a neutron star with $M_\mathrm{ns}=1.4\,M_\odot$. Both p and ns were assumed nonspinning and spherically symmetric. They were obtained by simultaneously integrating  the orbit-averaged equations for the de Sitter-only rate of change of ${\bds S}_\mathrm{p}$ \citep[][Equation\,(44) with Equation\,(45)]{1975PhRvD..12..329B} over $\Delta t = 1\,\mathrm{cty}$.  The initial conditions, corresponding to ${\bds{\hat{L}}},\,{\bds{\hat{S}}}_\mathrm{ns},\,{\bds{\hat{S}}}_\mathrm{p}$ arbitrarily oriented in space, are listed in Table\,\ref{table1}.
}\label{Netho_dS}
\end{figure}

\begin{figure}[H]
\centering
\centerline{
\vbox{
\begin{tabular}{c}
\epsfxsize= 16 cm\epsfbox{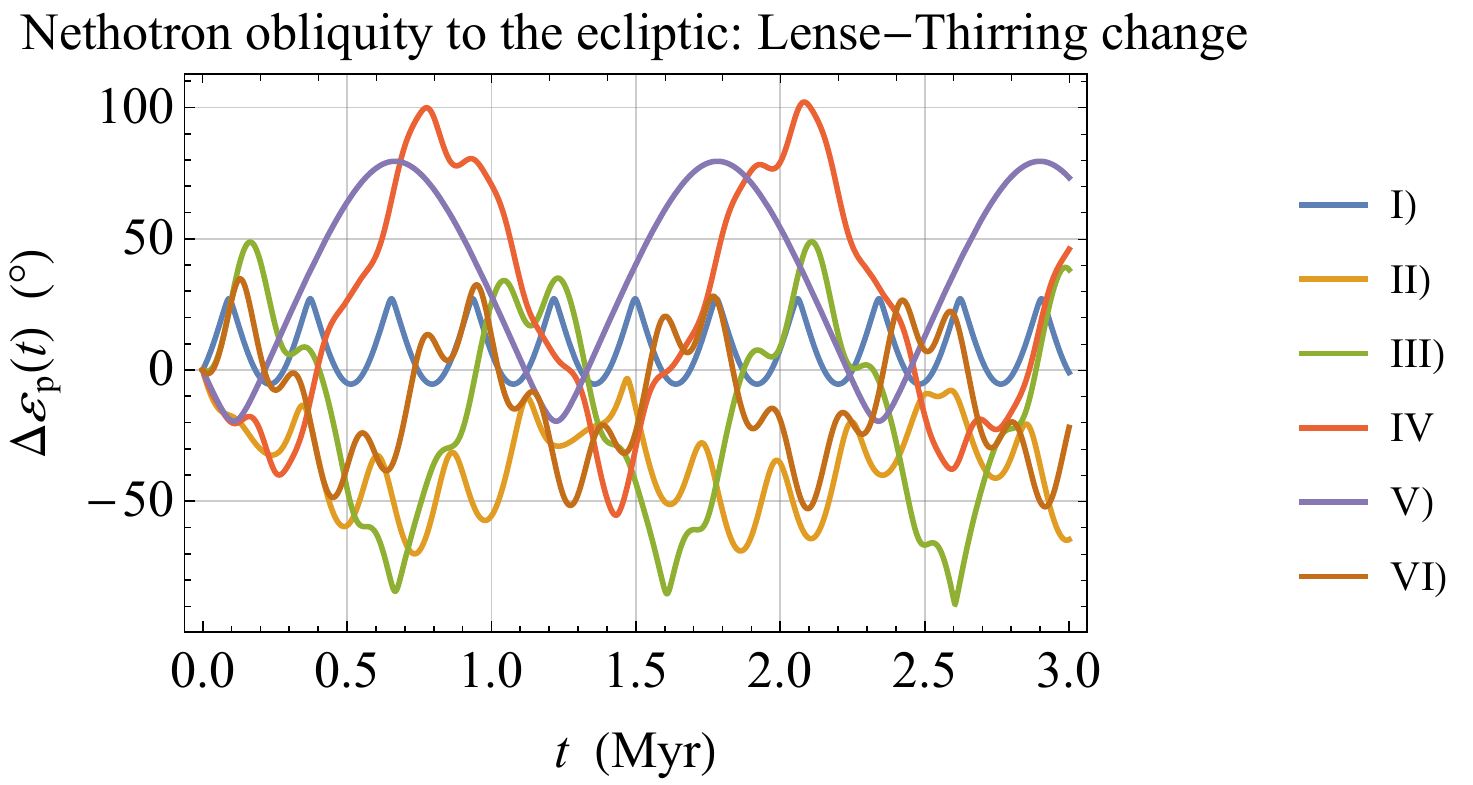}\\
\end{tabular}
}
}
\caption{
Numerically produced time series $\Delta \varepsilon_\mathrm{p}\ton{t} =  \varepsilon_\mathrm{p}\ton{t}- \varepsilon_\mathrm{p}^0$, in $^\circ$, of the pN gravitomagnetic variation of the obliquity $ \varepsilon_\mathrm{p}$ to the ecliptic plane of a putative Earth-like netohotron orbiting a neutron star with $M_\mathrm{ns}=1.4\,M_\odot$ and $S_\mathrm{ns}$ given by \rfr{spinNS}. The quadrupole mass moments $J_2^\mathrm{p},\,J_2^\mathrm{ns}$ of both p and ns  were set equal to zero. They were obtained by simultaneously integrating  the orbit-averaged equations for the Lense--Thirring-only rates of change of ${\bds S}_\mathrm{p}$ \citep[][Equation\,(44) with Equation\,(46)]{1975PhRvD..12..329B} and of $\bds L$ \citep[][Equation\,(64) with Equation\,(69)]{1975PhRvD..12..329B} over $\Delta t = 3\,\mathrm{Myr}$.  The initial conditions, corresponding to $\bds{\hat{L}},\,{{\bds{\hat{S}}}_\mathrm{p}}_\mathrm{ns},\,{{\bds{\hat{S}}}_\mathrm{p}}_\mathrm{p}$ arbitrarily oriented in space, are listed in Table\,\ref{table1}.
}\label{Netho_LT}
\end{figure}

\begin{figure}[H]
\centering
\centerline{
\vbox{
\begin{tabular}{c}
\epsfxsize= 16 cm\epsfbox{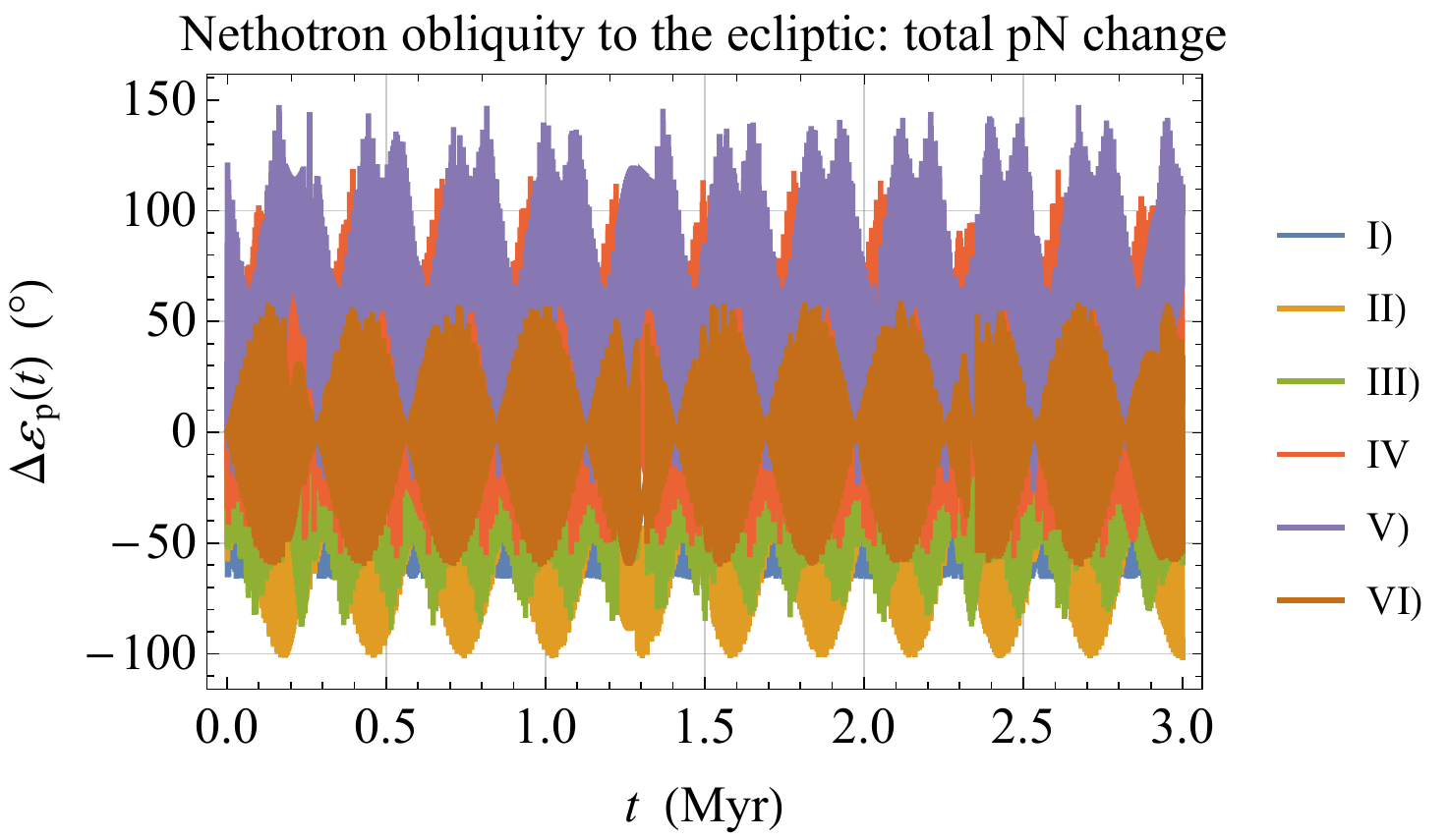}\\
\end{tabular}
}
}
\caption{
Numerically produced time series $\Delta \varepsilon_\mathrm{p}\ton{t} =  \varepsilon_\mathrm{p}\ton{t}- \varepsilon_\mathrm{p}^0$, in $^\circ$, of the total pN (de Sitter+ Lense--Thirring) variation of the obliquity $ \varepsilon_\mathrm{p}$ to the ecliptic plane of a putative Earth-like netohotron orbiting a neutron star with $M_\mathrm{ns}=1.4\,M_\odot$ and $S_\mathrm{ns}$ given by \rfr{spinNS}. They were obtained by simultaneously integrating  the orbit-averaged equations for the de Sitter and Lense--Thirring  rates of change of ${\bds S}_\mathrm{p}$ \citep[][Equation\,(44) with Equations\,(45)-(46)]{1975PhRvD..12..329B} and of $\bds L$ \citep[][Equation\,(64) with Equation\,(69)]{1975PhRvD..12..329B} over $\Delta t = 3\,\mathrm{Myr}$.  The initial conditions, corresponding to $\bds{\hat{L}},\,{{\bds{\hat{S}}}_\mathrm{p}}_\mathrm{ns},\,{{\bds{\hat{S}}}_\mathrm{p}}_\mathrm{p}$ arbitrarily oriented in space, are listed in Table\,\ref{table1}.
}\label{Netho_pN}
\end{figure}

\begin{figure}[H]
\centering
\centerline{
\vbox{
\begin{tabular}{c}
\epsfxsize= 16 cm\epsfbox{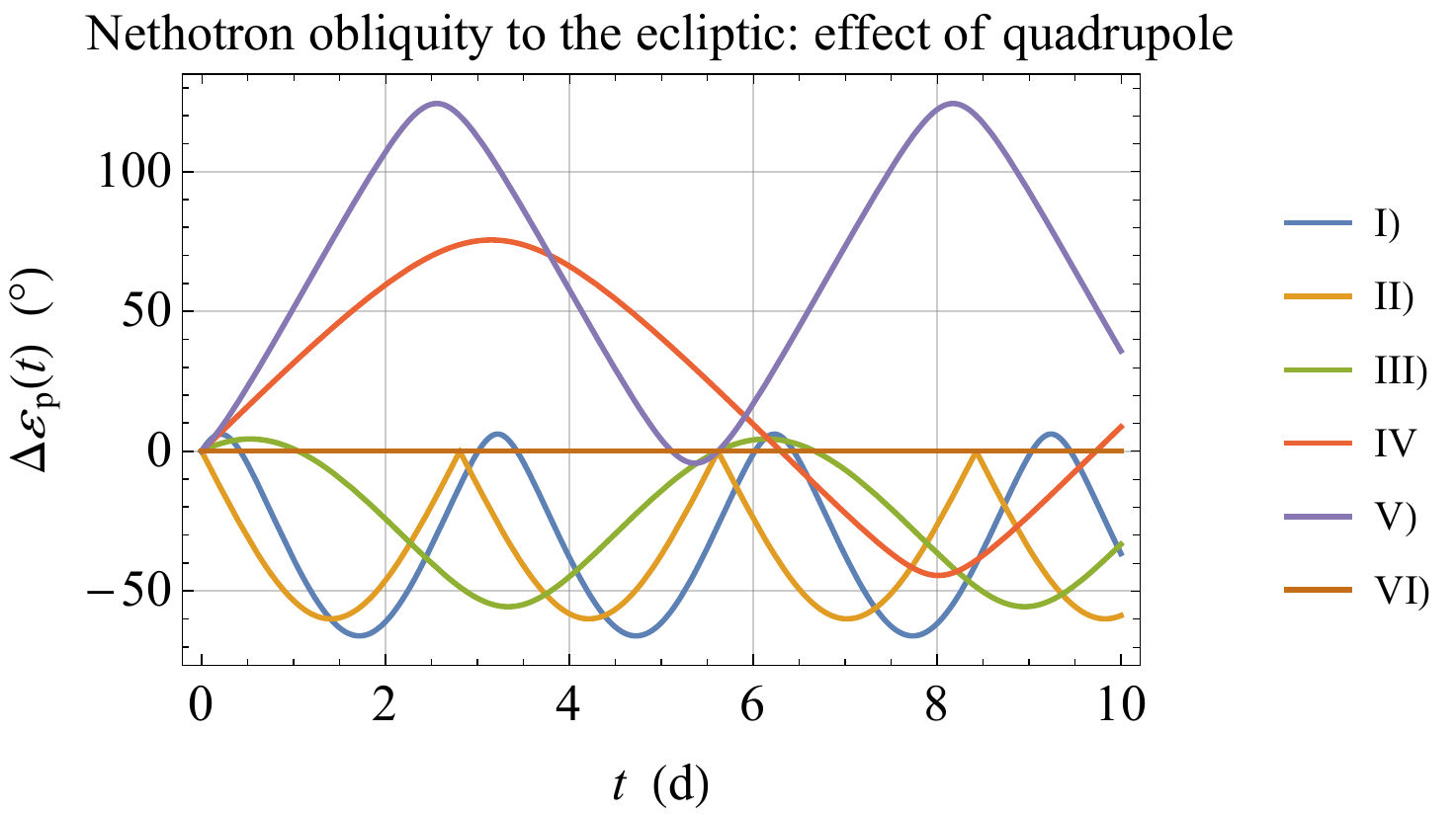}\\
\end{tabular}
}
}
\caption{
Numerically produced time series $\Delta \varepsilon_\mathrm{p}\ton{t} =  \varepsilon_\mathrm{p}\ton{t}- \varepsilon_\mathrm{p}^0$, in $^\circ$, of the classical variation of the obliquity $ \varepsilon_\mathrm{p}$ to the ecliptic plane of a putative oblate Earth-like netohotron orbiting a neutron star with $M_\mathrm{ns}=1.4\,M_\odot$, assumed spherical. They were obtained by simultaneously integrating  the orbit-averaged equations for the  Newtonian rates of change of ${\bds S}_\mathrm{p}$ \citep[][Equation\,(44) with Equation\,(47)]{1975PhRvD..12..329B} and of $\bds L$ \citep[][Equation\,(64) with Equation\,(71)]{1975PhRvD..12..329B} due to $J_2^\mathrm{p}$ over $\Delta t = 10\,\mathrm{d}$.  The initial conditions, corresponding to ${\bds{\hat{L}}},\,{\bds{\hat{S}}}_\mathrm{ns},\,{\bds{\hat{S}}}_\mathrm{p}$ arbitrarily oriented, are listed in Table\,\ref{table1}.
}\label{Netho_J2p}
\end{figure}

The initial conditions used in the runs are listed in Table\,\ref{table1}; they correspond to arbitrary initial spin-orbit configurations for
${\bds S}_\mathrm{ns},\,{\bds S}_\mathrm{p},$ and $\bds L$. In particular, their unit vectors at the initial epoch $t_0$ were parameterized as
\begin{align}
{\hat{S}}_x^\mathrm{ns}\ton{t_0} &=\sin\varepsilon_\mathrm{ns}^0\,\cos\alpha_\mathrm{ns}^0, \\ \nonumber\\
{\hat{S}}_y^\mathrm{ns}\ton{t_0} &=\sin\varepsilon_\mathrm{ns}^0\,\sin\alpha_\mathrm{ns}^0, \\ \nonumber\\
{\hat{S}}_z^\mathrm{ns}\ton{t_0} &=\cos\varepsilon_\mathrm{ns}^0, \\ \nonumber\\
{\hat{S}}_x^\mathrm{p}\ton{t_0} &=\sin \varepsilon_\mathrm{p}^0\,\cos\alpha_\mathrm{p}^0, \\ \nonumber\\
{\hat{S}}_y^\mathrm{p}\ton{t_0} &=\sin \varepsilon_\mathrm{p}^0\,\sin\alpha_\mathrm{p}^0, \\ \nonumber\\
{\hat{S}}_z^\mathrm{p}\ton{t_0} &=\cos \varepsilon_\mathrm{p}^0, \\ \nonumber\\
{\hat{L}}_x\ton{t_0} &=\sin I_0\,\sin\Omega_0, \\ \nonumber\\
{\hat{L}}_y\ton{t_0} &= -\sin I_0\,\cos\Omega_0, \\ \nonumber\\
{\hat{L}}_z\ton{t_0} &=\cos I_0, \\ \nonumber\\
{\hat{W}}_x\ton{t_0} \lb{Wx} &= 0, \\ \nonumber\\
{\hat{W}}_y\ton{t_0} &= 0, \\ \nonumber\\
{\hat{W}}_z\ton{t_0} \lb{Wz} &= 1,
\end{align}
where $\varepsilon_\mathrm{ns}^0,\,\alpha_\mathrm{ns}^0$ are the initial values of the obliquity to and of the azimuthal angle in the ecliptic plane of ${\bds{\hat{S}}}_\mathrm{ns}$, $ \varepsilon_\mathrm{p}^0,\,\alpha_\mathrm{p}^0$ are the initial values of the obliquity to and of the azimuthal angle in the ecliptic plane of ${\bds{\hat{S}}}_\mathrm{p}$, while $I_0$ and $\Omega_0$ are the initial values of the inclination and of the longitude of the ascending node of the nethotron's astrocentric orbital plane with respect to the ecliptic one. Note that the azimuthal angle $\Xi$ of ${\bds{\hat{L}}}$ is
\eqi
\Xi=\Omega-90^\circ.
\eqf
\begin{table}[!htb]
\begin{center}
\begin{threeparttable}
\caption{Initial Conditions Used in Figures\,\ref{Netho_dS}-\ref{Netho_J2p}.}
\label{table1}
\begin{tabular*}{\textwidth}{c@{\extracolsep{\fill}}c c c c c c c c c c}
\toprule
%
\multirow{2}{*}{}
& $\Pb$ (d) & $a$ ($\mathrm{au}$) & $e$ & $I_0$ ($^\circ$) & $\Omega_0$ ($^\circ$) & $\varepsilon_\mathrm{ns}^0$ ($^\circ$) & $\alpha_\mathrm{ns}^0$ ($^\circ$) & $ \varepsilon_\mathrm{p}^0$ ($^\circ$) & $\alpha_\mathrm{p}^0$ ($^\circ$) \\
\midrule
%
%
I)  & $0.109$   & $0.005$ & $0.0$ & $60$ & $0$  & $180$ & $300$ & $150$ & $120$  \\
II) & $0.109$   & $0.005$ & $0.0$ & $30$ & $60$ & $150$ & $240$ & $180$ & $300$   \\
III)& $0.109$  & $0.005$ & $0.0$ & $150$ & $120$ & $60$ & $180$ & $90$ & $240$  \\
IV) & $0.109$   & $0.005$ & $0.0$ & $120$ & $180$ & $90$ & $120$ & $60$ & $0$    \\
V)  & $0.109$   & $0.005$ & $0.0$ & $90$ & $240$ & $120$ & $60$ & $30$ & $180$   \\
VI) & $0.109$   & $0.005$ & $0.0$ & $180$ & $300$ & $30$ & $0$  & $120$ & $60$  \\
\bottomrule
\end{tabular*}
\begin{tablenotes}
\small
\item \textbf{Note.} Each row corresponds to the plotted times series with the same roman numeral in the legend of each figure. Recall that the azimuthal angle of ${\bds{\hat{L}}}$ is $\Xi=\Omega-90^\circ$.
\end{tablenotes}
\end{threeparttable}
\end{center}
\end{table}
I choose a close circular ($e=0.0$) orbit corresponding to the lowest value of the semimajor axis $a$ in the habitable zone of Figure\,5 in \citet{2017A&A...608A.147P}, i.e. $a=0.005\,\mathrm{au}$; thus, the orbital period amounts to
\eqi
\Pb=2.6\,\mathrm{hr}=0.109\,\mathrm{day}.
\eqf
The time series for the variation of the obliquity with respect to its initial value $\Delta \varepsilon_\mathrm{p}\ton{t}$ displayed in Figures\,\ref{Netho_dS}-\ref{Netho_J2p} were computed as
\eqi
\varepsilon_\mathrm{p}=\arccos\ton{{\bds{\hat{S}}}_\mathrm{p}\bds\cdot{\bds{\hat{W}}}}
\eqf
from the solutions for ${\hat{\bds S}}_\mathrm{p}$ and $\bds{\hat{W}}$. The $\grf{x,\,y}$ reference plane is the ecliptic plane at the initial epoch; cfr with \rfrs{Wx}{Wz}.

Figure\,\ref{Netho_dS} displays the numerically produced pN de Sitter-only obliquity variations $\Delta \varepsilon_\mathrm{p}$ of an Earth-like nethotron over $\Delta t= 1.4\,\mathrm{cty}$ induced by a neutron star with mass $M_\mathrm{ns}=1.4\,M_\odot$. Both p and ns are assumed nonspinning and spherically symmetric. The time series was obtained from the initial conditions of Table\,\ref{table1} by numerically integrating only Equation\,(44) with Equation\,(45) by \citet{1975PhRvD..12..329B} for the nethotron's spin ${\bds S}_\mathrm{p}$; the gravitoelectric static component of the pN field of a nonrotating body does not affect $\bds L$. It turns out that the characteristic time scale of all the signatures is $T\simeq 0.7\,\mathrm{cty}$, while their peak-to-peak amplitudes $ \varepsilon_\mathrm{p}^\mathrm{max}- \varepsilon_\mathrm{p}^\mathrm{min}$ can be as large as up to $\simeq 70^\circ-120^\circ$.

The numerically integrated pN gravitomagnetic obliquity changes for the same initial conditions of Table\,\ref{table1} are shown in Figure\,\ref{Netho_LT}. They were obtained by considering only the Lense--Thirring spin and orbital rates of change of the nethotron induced by the spin angular momentum ${\bds S}_\mathrm{ns}$ of the host neutron star calculated with \rfr{spinNS}. The integration interval is $\Delta t = 3\,\mathrm{Myr}$.
The temporal patterns turn out to be different from the purely gravitoelectric ones of Figure\,\ref{Netho_dS}, and the characteristic time scale is much longer, amounting to $T\simeq 1\,\mathrm{Myr}$. Also in this case, the peak-to-peak amplitudes of the signatures are relevant, reaching even $\simeq 50^\circ-100^\circ$.

Figure\,\ref{Netho_pN} displays the overall purely pN obliquity changes obtained by simultaneously integrating the equations for the rates of change of both the nethotron's spin and orbital angular momentum due to the pN gravitoeletric and gravitomagnetic field of the host neutron star: the quadrupole mass moments $J_2^\mathrm{ns},\,J_2^\mathrm{p}$ of both ns and p were set equal to zero. Also in this case, the integration interval is $\Delta t = 3\,\mathrm{Myr}$, and the initial conditions adopted are those in Table\,\ref{table1}. It can be noted that no mutual cancellation occur, amounting the amplitudes of $\Delta\varepsilon_\mathrm{p}$ to tens and even hundreds of degrees. It turns out that adding the classical shift of $\bds L$ due to the neutron star's quadrupole, calculated with \rfr{Qu2}, does not alter the pattern of Figure\,\ref{Netho_pN}.
%
%
%
In Figure\,\ref{Netho_J2p}, I show the purely Newtonian obliquity variations of the nethotron due to its own quadrupole mass moment $J_2^\mathrm{p}$ that displaces both its spin ${\bds S}_\mathrm{p}$ and its orbital angular momentum $\bds L$, as per Equation\,(44) with Equation\,(47) and Equation\,(64) with Equation\,(71) by \citet{1975PhRvD..12..329B}, respectively. The initial conditions are those in Table\,\ref{table1}, while the time span of the integration is $\Delta t=10\,\mathrm{days}$. The characteristic time is of the order of six days, and the peak-to-peak range of values $\varepsilon_\mathrm{p}^\mathrm{max}-\varepsilon_\mathrm{p}^\mathrm{min}$ can be as large as $\simeq 50^\circ-100^\circ$. Such high-frequency variations, which depend on the characteristic physical features of the nethotron itself, are to be thought of as superimposed to the much slower pN ones of Figure\,\ref{Netho_pN}.

In each of Figures\,\ref{Netho_dS}-\ref{Netho_J2p}, the impact of the very tiny and slow variations of the spin and orbital angular momenta induced by the distant massive light source S turn out to be negligible. For any practical purposes, the ecliptic plane can be considered fixed and the obliquity time series can be straightforwardly extracted from
\eqi
{\hat{S}}_z^\mathrm{p}\ton{t} = \cos\varepsilon_\mathrm{p}\ton{t}.
\eqf

About the particular scenario I adopted, I stress that it is just for illustrative purposes of the obliquity's phenomenology under consideration. At a close distance from its host neutron star, the nethotron would face huge tidal effects which may importantly contribute to its overall habitability. Indeed, the tidal dissipation rate in the planet scales as $a^{-1}\,\ton{R_\mathrm{p}/a}^{2\,\ell + 1}$, which, for $\ell=2$, becomes $\propto a^{-6}$ \citep[][Equation\,(65)]{2014ApJ...795....6E}. For such a tight orbit, the terms of degree even higher than $\ell=2$ may likely come into play as well. This indicates a powerful level of heating of tidal origin. Its quantitative assessment, which is outside the scope of this work, would be of relevance for the nethotron's ability to potentially sustain life one way or the other. If, on the one hand, the nethotron gets semimolten and
plastic for an excessive tidal overheating and, on the other hand, as pointed out by \citet{2017A&A...608A.147P}, tidal heating plays a positive role; the other sources of energy output from the neutron star should cease sooner or later.
The dynamical features of $\varepsilon_\mathrm{p}$ that I previously illustrated occur also for larger values of $a$ with longer characteristic timescales, while the tidal deformations would get less and less relevant.
\section{Summary and Conclusions}\lb{fine}
I looked at the possibility that potentially habitable rocky planets around neutron stars receive visible light from a distant point-like source like a main-sequence star at about 1 au or the gravitationally lensed accretion disk of a supermassive black hole at a distance such that it looks like our Sun as seen from Earth. In such a scenario, the planet's obliquity to the ecliptic plane is a key factor in assessing its habitability since it controls the insolation at a given latitude on its surface. Large and fast variations of the planet's axial tilt would have an impact on its long-term capability of hosting and sustaining life. It turned out that this may be just the case for rather general initial spin-orbit configurations because of the classical  torque exerted by the host neutron star on the planet's equatorial bulge and by the post-Newtonian de Sitter and Lense--Thirring  changes. In particular, for initially nonaligned spin and orbital angular momenta, the obliquity of an Earth-like rocky planet at, say, $0.005\,\mathrm{au}$ from a $1.4\,M_\odot$ neutron star, corresponding to the smallest admissible distance in the currently known habitable zone calculated for (isolated) neutron stars, experiences variations that may be as large as $\simeq 50^\circ-100^\circ$ over characteristic timescales ranging from about $10\,\mathrm{days}$ for the classical rate of change due to the planet's oblateness to $3\,\mathrm{Myr}$ for the post-Newtonian Lense--Thirring shift due to the neutron star's spin angular momentum. In view of such a distinctive feature, I suggested to name such planets as \virg{nethotrons.}

As directions for future work, I mention the need of revisiting the constraints on the radiative output from the host neutron star, on the magnetic and atmospheric environments of the planet when an additional external source of solar-like irradiation is present, and on the distance and the spectral class of the latter one in view of the effect of the neutron star itself in order to assure planetary conditions favorable to life. Furthermore, simulations aiming at determining the Galactic population of possible solar-neutron star binaries and the probability of their detection would be useful as well.
Finally, also the consequences of a potentially relevant tidal heating of the nethotron should be investigated in detail for tight orbits.

I thank  M. Efroimsky for remarks on the tidal heating  and A. Possenti and K. Postnov for useful information about possible binaries made of a neutron star and a Sun-type main-sequence star.
%
\bibliography{exopbib}{}

\begin{thebibliography}{31}
\expandafter\ifx\csname natexlab\endcsname\relax\def\natexlab#1{#1}\fi

\bibitem[{{Armstrong} {et~al}\mbox{.}(2014){Armstrong}, {Barnes},
  {Domagal-Goldman}, \& {et al.}}]{2014AsBio..14..277A}
{Armstrong} J.~C., {Barnes} R., {Domagal-Goldman} S., {et al.}, 2014, AsBio,
  14, 277

\bibitem[{{Bailes} {et~al}\mbox{.}(2011){Bailes}, {Bates}, {Bhalerao}, \& {et
  al.}}]{2011Sci...333.1717B}
{Bailes} M., {Bates} S.~D., {Bhalerao} V., {et al.}, 2011, Sci, 333, 1717

\bibitem[{{Barker} \& {O'Connell}(1975)}]{1975PhRvD..12..329B}
{Barker} B.~M., {O'Connell} R.~F., 1975, PhRvD, 12, 329

\bibitem[{{Bower} {et~al}\mbox{.}(2019){Bower}, {Chatterjee}, {Cordes}, \& {et
  al.}}]{2019BAAS...51c.438B}
{Bower} G., {Chatterjee} S., {Cordes} J., {et al.}, 2019, BAAS, 51, 438

\bibitem[{{Cordes} \& {Lazio}(1997)}]{1997ApJ...475..557C}
{Cordes} J.~M., {Lazio} T. J.~W., 1997, \apj, 475, 557

\bibitem[{{Correia} {et~al}\mbox{.}(2011){Correia}, {Laskar}, {Farago}, \&
  {Bou{\'e}}}]{2011CeMDA.111..105C}
{Correia} A. C.~M., {Laskar} J., {Farago} F., {Bou{\'e}} G., 2011, CeMDA, 111,
  105

\bibitem[{{Cushey}, {Majid} \& {Prince}(2017){Cushey}, {Majid}, \&
  {Prince}}]{2017AAS...22943103C}
{Cushey} D.~J., {Majid} W.~A., {Prince} T.~A., 2017, in 229th AAS Meeting
  Abstracts, (Washington: American Astronomical Society), p. 303

\bibitem[{{Deitrick} {et~al}\mbox{.}(2018){Deitrick}, {Barnes}, {Bitz}, \& {et
  al.}}]{2018AJ....155..266D}
{Deitrick} R., {Barnes} R., {Bitz} C., {et al.}, 2018, \aj, 155, 266

\bibitem[{{Efroimsky} \& {Makarov}(2014)}]{2014ApJ...795....6E}
{Efroimsky} M., {Makarov} V.~V., 2014, \apj, 795, 6

\bibitem[{{Genzel}, {Eisenhauer} \& {Gillessen}(2010){Genzel}, {Eisenhauer}, \&
  {Gillessen}}]{2010RvMP...82.3121G}
{Genzel} R., {Eisenhauer} F., {Gillessen} S., 2010, RvMP, 82, 3121

\bibitem[{{Greaves} \& {Holland}(2000)}]{2000MNRAS.316L..21G}
{Greaves} J.~S., {Holland} W.~S., 2000, \mnras, 316, L21

\bibitem[{{Haghighipour}(2013)}]{2013AREPS..41..469H}
{Haghighipour} N., 2013, AREPS, 41, 469

\bibitem[{{Haghighipour}(2015)}]{2015enas.book.2436H}
{Haghighipour} N., 2015, in Encyclopedia of Astrobiology, {Gargaud} M.,
  {Irvine} W.~M., {Amils} R., {et al.}, eds., (Berlin, Heidelberg: Springer),
  pp. 2436--2436

\bibitem[{{H{\'e}brard} {et~al}\mbox{.}(2004){H{\'e}brard}, {Lecavelier Des
  {\'E}tangs}, {Vidal-Madjar}, {D{\'e}sert}, \& {Ferlet}}]{2004ASPC..321..203H}
{H{\'e}brard} G., {Lecavelier Des {\'E}tangs} A., {Vidal-Madjar} A.,
  {D{\'e}sert} J.~M., {Ferlet} R., 2004, in ASP Conference Series, Vol. 321,
  Extrasolar Planets: Today and Tomorrow, {Beaulieu} J., {Lecavelier Des
  Etangs} A., {Terquem} C., eds., (Orem, UT: Astronomical Society of the
  Pacific), pp. 203--204

\bibitem[{{Kim} \& {Davies}(2018)}]{2018JKAS...51..165K}
{Kim} C., {Davies} M.~B., 2018, JKAS, 51, 165

\bibitem[{{Morris} \& {Serabyn}(1996)}]{1996ARA&A..34..645M}
{Morris} M., {Serabyn} E., 1996, ARA\&A, 34, 645

\bibitem[{{Patruno} \& {Kama}(2017)}]{2017A&A...608A.147P}
{Patruno} A., {Kama} M., 2017, A\&A, 608, A147

\bibitem[{{Podsiadlowski}(1995)}]{1995ASPC...72..411P}
{Podsiadlowski} P., 1995, in ASP Conference Series, Vol.~72, Millisecond
  Pulsars. A Decade of Surprise, {Fruchter} A.~S., {Tavani} M., {Backer} D.~C.,
  eds., (Orem, UT: Astronomical Society of the Pacific), pp. 411--420

\bibitem[{{Poisson} \& {Will}(2014)}]{2014grav.book.....P}
{Poisson} E., {Will} C.~M., 2014, {Gravity}. (Cambridge: Cambridge Univ. Press)

\bibitem[{{Rajwade}, {Lorimer} \& {Anderson}(2017){Rajwade}, {Lorimer}, \&
  {Anderson}}]{2017MNRAS.471..730R}
{Rajwade} K.~M., {Lorimer} D.~R., {Anderson} L.~D., 2017, \mnras, 471, 730

\bibitem[{{Schnittman}(2019)}]{2019arXiv191000940S}
{Schnittman} J.~D., 2019, arXiv:1910.00940

\bibitem[{{Silva} {et~al}\mbox{.}(2021){Silva}, {Holgado},
  {C{\'a}rdenas-Avenda{\~n}o}, \& {Yunes}}]{2021PhRvL.126r1101S}
{Silva} H.~O., {Holgado} A.~M., {C{\'a}rdenas-Avenda{\~n}o} A., {Yunes} N.,
  2021, PhRvL, 126, 181101

\bibitem[{{Spiegel}, {Menou} \& {Scharf}(2009){Spiegel}, {Menou}, \&
  {Scharf}}]{2009ApJ...691..596S}
{Spiegel} D.~S., {Menou} K., {Scharf} C.~A., 2009, \apj, 691, 596

\bibitem[{{Wada}, {Tsukamoto} \& {Kokubo}(2019){Wada}, {Tsukamoto}, \&
  {Kokubo}}]{2019ApJ...886..107W}
{Wada} K., {Tsukamoto} Y., {Kokubo} E., 2019, \apj, 886, 107

\bibitem[{{Wada}, {Tsukamoto} \& {Kokubo}(2021{\natexlab{a}}){Wada},
  {Tsukamoto}, \& {Kokubo}}]{2021ApJ...907...58W}
{Wada} K., {Tsukamoto} Y., {Kokubo} E., 2021{\natexlab{a}}, \apj, 907, 58

\bibitem[{{Wada}, {Tsukamoto} \& {Kokubo}(2021{\natexlab{b}}){Wada},
  {Tsukamoto}, \& {Kokubo}}]{2021ApJ...909...96W}
{Wada} K., {Tsukamoto} Y., {Kokubo} E., 2021{\natexlab{b}}, \apj, 909, 96

\bibitem[{{Wang}(2014)}]{2014P&SS..100...19W}
{Wang} Z., 2014, P\&SS, 100, 19

\bibitem[{{Wex}(1998)}]{1998MNRAS.298...67W}
{Wex} N., 1998, \mnras, 298, 67

\bibitem[{{Wolszczan}(1994)}]{1994Sci...264..538W}
{Wolszczan} A., 1994, Sci, 264, 538

\bibitem[{{Wolszczan}(2006)}]{2006ISSIR...6...77W}
{Wolszczan} A., 2006, ISSIR, 6, 77

\bibitem[{{Wolszczan} \& {Frail}(1992)}]{1992Natur.355..145W}
{Wolszczan} A., {Frail} D.~A., 1992, Natur, 355, 145

\end{thebibliography}

\end{document}